\documentstyle[aps,pra]{revtex}
\begin{document}
\title{Multi-directional higher-order amplitude squeezing}
\author{Nguyen Ba An\thanks{%
Permanent address: Institute of Physics, P. O. Box 429 Bo Ho, Hanoi 10000,
Vietnam. Email: nbaan@netnam.org.vn.}}
\address{National Center for Theoretical Sciences, Physics Division, P. O. Box 2-131, 
\\
Hsinchu, Taiwan 300, R. O. C. }
\maketitle

\begin{abstract}
Fan-even K-quantum nonlinear coherent states are introduced and higher-order
amplitude squeezing is investigated in such states. It is shown that for a
given $K$ the lowest order in which an amplitude component can be squeezed
is $2K$ and the squeezing appears simultaneously in $K$ directions separated
successively in phase by $\pi /K.$

PACS numbers: 42.50.Dv
\end{abstract}

\pacs{PACS numbers: 42.50.Dv}

\noindent {\bf 1. Introduction}

During a few last years, there has been growing interest in the nonlinear
coherent state (NCS) \cite{ncs1,ncs2,ncs3,ncs4,ncs5} defined as the
eigen-state $\left| \chi ;f\right\rangle $ of the non-boson non-Hermitian
operator $af(\widehat{n}),$%
\begin{equation}
af(\widehat{n})\left| \chi ;f\right\rangle =\chi \left| \chi ;f\right\rangle
,  \label{1}
\end{equation}
with $\widehat{n}=a^{+}a,$ $a$ the bosonic annihilation operator, $\chi $ a
complex eigen-value and $f$ an arbitrary (assumed to be real) nonlinear
operator-valued function of $\widehat{n}.$ Recently, the so called K-quantum
nonlinear coherent state (KNCS) has been introduced \cite{5a,5b,an} as a
generalization of the NCS to $K$ ($K$ an arbitrary positive integer)
eigen-states $\left| \xi ;K,j,f\right\rangle $ of the non-boson
non-Hermitian operator $a^{K}f(\widehat{n}),$ 
\begin{equation}
a^{K}f(\widehat{n})\left| \xi ;K,j,f\right\rangle =\xi ^{K}\left| \xi
;K,j,f\right\rangle ,  \label{2}
\end{equation}
with $j=0,1,...,K-1$ and $\xi $ a complex number. Note that the notations in 
\cite{5a,5b,an} are not the same. In \cite{5a} $(af(\widehat{n}))^{k}$ is
used instead of $a^{K}f(\widehat{n}).$ In \cite{5b} it is $a^{N+1}f(\widehat{%
n})$ with $N$ counted from zero and the state is named NCS of order $N+1.$
The even (odd) NCS \cite{oe1,oe2,oe3} corresponds to $K=2,$ $j=0$ $(j=1).$
The linear case (see, e.g. \cite{l1,l2,l3,l4,l5,l6,l7}) is recovered when $%
f\equiv 1.$ Other types of multi-quantum states are also developed (see,
e.g., \cite{k1,k2}). In \cite{5b,an} the KNCS has been shown physically
realizable in the quantized vibration of the center-of-mass motion of a
harmonically trapped ion which is further properly driven by two laser beams
one of which is resonant and the other is detuned to the $K$th lower
sideband. The KNCS displays a multi-peaked structure in the number
distribution depending on both the character of the nonlinear function $f$
and the power $K.$ The normalization difficulty connected with the wildly
oscillating behavior at large $n$ and Lamb-Dicke parameter has been dealt
with in \cite{5b} while nonclassical effects have been studied in \cite{an}.
In this work we further explore the KNCS under another angle of view.
Namely, a property of the KNCS will be used to construct the so-called
fan-even K-quantum nonlinear coherent states (FEKNCS's) in which
higher-order amplitude squeezing will be investigated. Unlike in usual
states, in FEKNCS's multi-directional squeezing is possible.

\noindent {\bf 2. Fan-even K-quantum nonlinear coherent state}

Spanned in Fock space the KNCS has the explicit form 
\begin{equation}
\left| \xi ;K,j,f\right\rangle =C_{Kj}\sum_{n=0}^{\infty }\frac{\xi ^{nK+j}}{%
\sqrt{(nK+j)!}f(nK+j)!}\left| nK+j\right\rangle  \label{8}
\end{equation}
where 
\begin{equation}
f(nK+j)!\equiv \left\{ 
\begin{array}{ll}
\prod_{q=0}^{n}f(qK+j), & n\geq 1 \\ 
1, & n=0
\end{array}
\right.  \label{4}
\end{equation}
and the coefficient $C_{Kj}$ is determined as 
\begin{equation}
C_{Kj}\equiv C_{Kj}(|\xi |^{2})=\left[ \sum_{m=0}^{\infty }\frac{|\xi
|^{2(mK+j)}}{(mK+j)!\left[ f(mK+j)!\right] ^{2}}\right] ^{-1/2}  \label{7}
\end{equation}
in order to normalize the KNCS to $1.$ Among the various properties of the
KNCS (see \cite{5a,5b,an}), the one of our direct concern here is that any
state $\left| \xi ;K,j,f\right\rangle $ can be decomposed into a correlated
combination of $K$ states $\left| \chi _{l};f\right\rangle $ as 
\begin{equation}
\left| \xi ;K,j,f\right\rangle =\frac{1}{K}\frac{C_{Kj}(|\xi |^{2})}{%
C_{10}(|\xi |^{2})}\sum_{l=0}^{K-1}\exp \left( -\frac{2\pi i}{K}jl\right)
\left| \chi _{l};f\right\rangle  \label{13}
\end{equation}
with $\left| \chi _{l};f\right\rangle $ the NCS defined by Eq. (\ref{1}) and 
\begin{equation}
\chi _{l}=\xi \exp \left( \frac{2\pi i}{K}l\right) .  \label{14a}
\end{equation}
The decomposition (\ref{13}) can be verified straightforwardly by
substituting $\left| \chi _{l};f\right\rangle \equiv \left| \chi
_{l};1,0,f\right\rangle $ and (\ref{14a}) into the r.h.s. of (\ref{13}) with
subsequent use of the identity 
\begin{equation}
\sum_{l=0}^{L-1}\exp \left[ \frac{2\pi i}{L}ql\right] \equiv L\delta _{q,mL}%
\text{ \quad with }m\text{ a nonnegative integer.}
\end{equation}
Equation (\ref{13}) represents a superposition state belonging to the type
of generalized NCS 
\begin{equation}
\left| \Psi _{K}\right\rangle =\sum_{l=0}^{K-1}a_{l}^{(K)}\left| \xi \exp
(i\alpha _{l}^{(K)});f\right\rangle
\end{equation}
whose linear version with $f\equiv 1$ was found in \cite{g1} and further
developed in \cite{g2,g3} (see also \cite{b1,b2} and references therein for
general types of superposition states, still with $f\equiv 1).$ Generally,
by choosing suitably the weights $a_{l}^{(K)}$ and phases $\alpha _{l}^{(K)}$
the state $\left| \Psi _{K}\right\rangle $ can be tailored for various kinds
of state. The choice $\alpha _{l}^{(K)}=2\pi l/K$ and $a_{l}^{(K)}=C_{Kj}%
\exp (-2\pi ijl/K)/(KC_{10})$ fits $\left| \Psi _{K}\right\rangle $ to be
the normalized KNCS $\left| \xi ;K,j,f\right\rangle .$ The ends of $\chi
_{l},$ Eq. (\ref{14a}), are equi-distantly spaced along a circle of radius $%
|\xi |$ in the complex plane and $\left| \xi ;K,j,f\right\rangle $ can be
referred to as state on a circle \cite{jan} or circular state \cite{ragi}.

Let $T_{m}$ be the rotation operator that rotates the $\chi _{l}$ on an
angle $\varphi _{m}=2\pi m/K$ ($m=0,1,...,K-1),$ i.e. 
\[
T_{m}\left| \chi _{l};f\right\rangle =\left| \chi _{l}^{\prime
};f\right\rangle =\left| \exp \left( \frac{2\pi i}{K}m\right) \chi
_{l};f\right\rangle . 
\]
Under such a rotation the KNCS gains a $j$-dependent phase shift 
\begin{equation}
T_{m}\left| \xi ;K,j,f\right\rangle =\exp \left( \frac{2\pi i}{K}jm\right)
\left| \xi ;K,j,f\right\rangle  \label{t}
\end{equation}
as is evident from (\ref{13}). The transformation (\ref{t}) indicates that
in general the KNCS $\left| \xi ;K,j,f\right\rangle $ cannot be identified
as even or odd in the usual sense with respect to the inversion $\xi
\rightarrow -\xi $ corresponding a $\pi $-rotation. More precisely, for odd $%
K$ the KNCS is neither even nor odd. However, for even $K$ the states $%
\left| \xi ;K,j,f\right\rangle $ may be either even or odd depending on the
evenness of $j.$ If $j$ is even (odd) the KNCS is even (odd) too. Moreover,
when $j=0$ the states $\left| \xi ;K,0,f\right\rangle $ for any $K$ turn out
to be symmetric in the sense of their invariance under the rotation $T_{m}$
with any $m=0,1,...,K-1,$ as is transparent from (\ref{t}). If, in addition
to $j=0,$ $K$ is even then the states become at the same time both symmetric
and even. In what follows we shall need such symmetric-even K-quantum
nonlinear coherent states (SEKNCS's), i.e. the states $\left| \xi
;K,0,f\right\rangle $ with even $K,$ and denote them by $\left| \xi
;K,f\right\rangle _{se},$ 
\begin{equation}
\left| \xi ;K,f\right\rangle _{se}=\frac{1}{K}\frac{C_{K0}(|\xi |^{2})}{%
C_{10}(|\xi |^{2})}\sum_{l=0}^{K-1}\left| \chi _{l};f\right\rangle \text{
with }K\text{ even,}  \label{se}
\end{equation}
where the subscript ``$se"$ stands for simultaneous ``symmetric'' and
``even''. ``Simultaneous'' is crucial because the KNCS may be symmetric but
not even (e.g., for $j=0,$ $K$ odd) or even but asymmetric (e.g., $K$ even, $%
j=2,4,...,K-2\neq 0$). In particular, when $K=2$ the states $\left| \xi
;2,f\right\rangle _{se}$ are simply referred to as even nonlinear coherent
states since in this special case ``symmetric'' and ``even'' coincide.
Figure 1a represents the orientation of the $\chi _{l}$ in the complex plane
in a SEKNCS with $K=8.$ For an odd $K$ the $\chi _{l}$ point symmetrically
too but the state made of them is neither even nor odd (see Fig. 1b, for $%
K=7).$

We now construct a state denoted by $\left| \xi ;K,f\right\rangle _{F}$
which is superposed by SEKNCS's $\left| \xi _{q};K,f\right\rangle _{se}$ in
the following way 
\begin{equation}
\left| \xi ;K,f\right\rangle _{F}=B_{K}\sum_{q=0}^{K-1}\left| \xi
_{q};K,f\right\rangle _{se}.  \label{fan}
\end{equation}
In the definition (\ref{fan}) 
\begin{equation}
\xi _{q}=\xi \exp \left( \frac{\pi i}{K}q\right)  \label{xiq}
\end{equation}
and $B_{K}\equiv B_{K}(|\xi |^{2})$ the normalization coefficient determined
from the equation 
\begin{equation}
B_{K}^{2}(|\xi |^{2})C_{K0}^{2}(|\xi |^{2})D_{K}(|\xi |^{2})=1  \label{b}
\end{equation}
where 
\begin{equation}
D_{K}(|\xi |^{2})=\sum_{m=0}^{\infty }\frac{\left| \xi \right| ^{2mK}\left|
J_{K}(m)\right| ^{2}}{(mK)!\left[ f(mK)!\right] ^{2}}  \label{d}
\end{equation}
with $J_{K}(m)$ given by 
\begin{equation}
J_{K}(m)=\sum_{q=0}^{K-1}\exp (i\pi qm).  \label{J}
\end{equation}
The state $\left| \xi ;K,f\right\rangle _{F}$ defined by Eq. (\ref{fan}) is
in fact a K-quantum superposition state made of the component states $\left|
\xi _{q};K,f\right\rangle _{se}$ which is also a K-quantum superposition
state over the simpler single-quantum states $\left| \chi
_{l};f\right\rangle \equiv \left| \chi _{l};1,f\right\rangle .$ In this
respect $\left| \xi _{q};K,f\right\rangle _{se}$ can be considered as the
primary superposition states and $\left| \xi ;K,f\right\rangle _{F}$ the
secondary ones. This makes sense toward a production scheme: the primary
superposition states should be generated first from the single-quantum
states and then, in the second step, these are to be used as the input for
the secondary superposition state as the output. The orientation of $\xi
_{q},$ Eq. (\ref{xiq}), in the complex plane (Fig. 2) looks like an open
paper fan. Hence we call the state $\left| \xi ;K,f\right\rangle _{F}$
fan-even K-quantum nonlinear coherent states (FEKNCS's) with the subscript ``%
$F"$ standing for ``fan''. In what follows FEKNCS's will be referred in
short to as fan-states. When $K=2$ the fan shrinks to a setsquare ({\it une} 
{\it equerre}). That is why the state 
\begin{equation}
\left| \xi ;2,f\right\rangle _{F}=B_{2}\left( \left| \xi ;2,f\right\rangle
_{se}+\left| i\xi ;2,f\right\rangle _{se}\right)  \label{f2}
\end{equation}
was named orthogonal-even nonlinear coherent state \cite{das} which is the
simplest fan-state. In addition, if $f\equiv 1,$ the state (\ref{f2})
reduces to that proposed in \cite{lynch}.

\noindent {\bf 3. Multi-directional higher-order amplitude squeezing}

Consider a boson field with the annihilation and creation operators $a$ and $%
a^{+}$ obeying the Bose-Einstein commutation relation $[a,a^{+}]=1.$ Let $%
X_{\varphi }$ be a field amplitude component pointing along the direction
making an angle $\varphi $ with the real axis in the complex plane 
\begin{equation}
X_{\varphi }=(a\text{ e}^{-i\varphi }+a^{+}\text{ e}^{i\varphi })/\sqrt{2}.
\label{am}
\end{equation}
The $\sqrt{2}$ was used above instead of $2$ in the definition of $%
X_{\varphi }$ is just a matter of notation. A state $\left| ...\right\rangle 
$ is said to be amplitude-squeezed to the $N^{th}$ order ($N=2,4,6,...)$
along the direction $\varphi $ if the quantity $S_{\varphi ,N},$%
\begin{equation}
S_{\varphi ,N}=\left\langle (\Delta X_{\varphi })^{N}\right\rangle -\left(
N-1\right) !!/\sqrt{2^{N}},  \label{Sn}
\end{equation}
with $\Delta X_{\varphi }\equiv X_{\varphi }-\left\langle X_{\varphi
}\right\rangle ,$ gets negative \cite{HongMandel}.

Let the real axis be chosen along the direction of $\xi .$ This allows
treating $\xi $ as a real number. We now proceed to study in detail various
higher-order amplitude squeezings in the above constructed fan-states $%
\left| \xi ;K,f\right\rangle _{F}.$ The problem depends essentially on the
concrete form of the nonlinear function $f(\widehat{n})$ which differs
strongly from one to another physical context (e.g., see \cite{harmonious}
for harmonious oscillators, \cite{pacs} for photon-added coherent states, 
\cite{ncs1,an,ion} for trapped ions, etc.) and requires formidable numerical
simulations. Nevertheless, it is worth stressing that our main target here
is to show a possible multi-directional character of squeezing in the
fan-state. Since $f$ depends only on $\widehat{n},$ i.e. it is
phase-independent, the squeezing directions are not affected by $f.$ Its
inclusion due to its nonlinearity will only modify the parameter space where
squeezing may appear. By this reason and to pursue our primary goal we limit
ourselves in this work to $f\equiv 1$ which allows us to achieve the general
results at a fully analytic level. The power $K$ is however kept arbitrary.

For $K=N=2$ we have obtained 
\begin{equation}
S_{\varphi ,2}^{(K=2)}=\frac{\xi ^{2}}{D_{2}}\left[ \sinh (\xi ^{2})-\sin
(\xi ^{2})\right]  \label{s122}
\end{equation}
with 
\begin{equation}
D_{2}=\cosh (\xi ^{2})+\cos (\xi ^{2}).  \label{D2}
\end{equation}
Since $S_{\varphi ,2}^{(K=2)}$ is independent of $\varphi $ and always
positive, the conventional amplitude squeezing ($N=2$) is totally absent.

For $K=2$ and $N=4$ we have obtained 
\begin{equation}
S_{\varphi ,4}^{(K=2)}=\frac{\xi ^{2}}{2}\left\{ \xi ^{2}\cos (4\varphi )-%
\frac{3}{D_{2}}\left[ \xi ^{2}\left( \cos (\xi ^{2})-\cosh (\xi ^{2})\right)
+2\left( \sin (\xi ^{2})-\sinh (\xi ^{2})\right) \right] \right\} .
\label{s124}
\end{equation}
It follows from (\ref{s124}) that the $4^{th}$ order squeezing occurs
whenever 
\begin{equation}
\cos (4\varphi )<g(|\xi |)=\frac{3\left[ \xi ^{2}\left( \cos (\xi
^{2})-\cosh (\xi ^{2})\right) +2\left( \sin (\xi ^{2})-\sinh (\xi
^{2})\right) \right] }{\xi ^{2}\left( \cosh (\xi ^{2})+\cos (\xi
^{2})\right) }.  \label{g124}
\end{equation}
The $g(|\xi |)$ equals zero at $\xi =0.$ As $|\xi |$ increases, $g$ first
decreases then, after reaching a minimum, increases up and tends to $-3$
when $|\xi |\rightarrow \infty .$ No squeezing appears for $|\xi |\geq \xi
_{c}=0.796541$ for which $g(|\xi |)\leq -1$ and no $\varphi $ can be found
to make $S_{\varphi ,4}^{(K=2)}$ negative. Yet, for $0<|\xi |<\xi _{c}$
there exist intervals of $\varphi $ that fulfils the condition (\ref{g124}).
Let us define the squeezing (stretching) direction as that along which the
amplitude is maximally squeezed (unsqueezed). Then it can be verified that
the squeezing directions are along $\varphi =\varphi _{sq,1}=\pi /4$ $(5\pi
/4)$ and $\varphi =\varphi _{sq,2}=3\pi /4$ $(7\pi /4)$ while for the
stretching directions $\varphi =\varphi _{st,1}=0$ $(\pi )$ and $\varphi
=\varphi _{st,2}=\pi /2$ $(3\pi /2).$ Along a squeezing direction a maximal
squeezing is reached at $|\xi |=\xi _{M}=0.669272.$ The existence of two
squeezing directions that are orthogonal to each other is owing to the
symmetry of the fan-state under consideration with $K=2.$

For $K=2$ and $N=6$ we have obtained 
\begin{eqnarray}
S_{\varphi ,6}^{(K=2)} &=&\frac{\xi ^{2}}{2}\left\{ \left[ \frac{15}{2}+%
\frac{3\xi ^{2}}{D_{2}}\left( \sinh (\xi ^{2})-\sin (\xi ^{2})\right)
\right] \cos \left( 4\varphi \right) \right.  \nonumber \\
&&+\frac{1}{2D_{2}}\left[ 10\xi ^{4}\left( \sinh (\xi ^{2})+\sin (\xi
^{2})\right) \right.  \nonumber \\
&&\left. \left. +45\xi ^{2}\left( \cosh (\xi ^{2})-\cos (\xi ^{2})\right)
+45\left( \sinh (\xi ^{2})-\sin (\xi ^{2})\right) \right] \right\} .
\end{eqnarray}
Figure 3a is a 3D plot of $S_{\varphi ,6}^{(K=2)}$ as a function of $|\xi |$
and $\varphi $ showing again maximal squeezing (stretching) occurred along
the directions $\varphi =\varphi _{sq,1}$ and $\varphi =\varphi _{sq,2}$ ($%
\varphi =\varphi _{st,1}$ and $\varphi =\varphi _{st,2})$. The values of $%
|\xi |$ for which squeezing appears are confined within the interval $%
[0,0.785486]$ and a maximal squeezing is gained at $|\xi |=0.659675.$ The
two coexistent directions of squeezing are most visualized by a polar plot
(Fig. 3b) in which the uncertainty circle of radius $r=15/8$ for the
coherent state is deformed into a four-winged flower. More flower-like is
the $S_{\varphi ,6}^{(K=2)}$ itself when it is watched upon in polar
coordinates (Fig. 3c).

For $K=4$ and $N=2,4,6$ we have obtained 
\begin{eqnarray}
S_{\varphi ,2}^{(K=4)} &=&\frac{\xi ^{2}}{D_{4}}\left\{ \sinh (\xi
^{2})-\sin (\xi ^{2})\right.  \nonumber \\
&&\left. +\sqrt{2}\left[ \sinh \left( \frac{\xi ^{2}}{\sqrt{2}}\right) \cos
\left( \frac{\xi ^{2}}{\sqrt{2}}\right) -\sin \left( \frac{\xi ^{2}}{\sqrt{2}%
}\right) \cosh \left( \frac{\xi ^{2}}{\sqrt{2}}\right) \right] \right\} ,
\end{eqnarray}
\begin{eqnarray}
S_{\varphi ,4}^{(K=4)} &=&\frac{3\xi ^{2}}{2D_{4}}\left\{ \xi ^{2}\left[
\cosh (\xi ^{2})-\cos (\xi ^{2})-2\sinh \left( \frac{\xi ^{2}}{\sqrt{2}}%
\right) \sin \left( \frac{\xi ^{2}}{\sqrt{2}}\right) \right] \right. 
\nonumber \\
&&+\sinh (\xi ^{2})-\sin (\xi ^{2})  \nonumber \\
&&\left. +\sqrt{2}\left[ \sinh \left( \frac{\xi ^{2}}{\sqrt{2}}\right) \cos
\left( \frac{\xi ^{2}}{\sqrt{2}}\right) -\sin \left( \frac{\xi ^{2}}{\sqrt{2}%
}\right) \cosh \left( \frac{\xi ^{2}}{\sqrt{2}}\right) \right] \right\} ,
\end{eqnarray}
\begin{eqnarray}
S_{\varphi ,6}^{(K=4)} &=&\frac{5\xi ^{2}}{4D_{4}}\left\{ 2\xi ^{4}\left[
\sinh (\xi ^{2})+\sin (\xi ^{2})\right. \right.  \nonumber \\
&&\left. -\sqrt{2}\left[ \sinh \left( \frac{\xi ^{2}}{\sqrt{2}}\right) \cos
\left( \frac{\xi ^{2}}{\sqrt{2}}\right) +\sin \left( \frac{\xi ^{2}}{\sqrt{2}%
}\right) \cosh \left( \frac{\xi ^{2}}{\sqrt{2}}\right) \right] \right] 
\nonumber \\
&&+9\xi ^{2}\left[ \cosh (\xi ^{2})-\cos (\xi ^{2})-2\sinh \left( \frac{\xi
^{2}}{\sqrt{2}}\right) \sin \left( \frac{\xi ^{2}}{\sqrt{2}}\right) \right] 
\nonumber \\
&&+9\left[ \sinh (\xi ^{2})-\sin (\xi ^{2})\right.  \nonumber \\
&&\left. \left. +\sqrt{2}\left[ \sinh \left( \frac{\xi ^{2}}{\sqrt{2}}%
\right) \cos \left( \frac{\xi ^{2}}{\sqrt{2}}\right) -\sin \left( \frac{\xi
^{2}}{\sqrt{2}}\right) \cosh \left( \frac{\xi ^{2}}{\sqrt{2}}\right) \right]
\right] \right\} .
\end{eqnarray}
with 
\begin{equation}
D_{4}=\cosh (\xi ^{2})+\cos (\xi ^{2})+2\cos \left( \frac{\xi ^{2}}{\sqrt{2}}%
\right) \cosh \left( \frac{\xi ^{2}}{\sqrt{2}}\right)
\end{equation}
The $S_{\varphi ,2(4,6)}^{(K=4)}$ are all independent of $\varphi $ and
always positive resulting in no squeezing.

For $K=4$ and $N=8$ we have obtained 
\begin{eqnarray}
S_{\varphi ,8}^{(K=4)} &=&\frac{\xi ^{2}}{8}\left\{ \xi ^{6}\cos (8\varphi )+%
\frac{1}{D_{4}}\left[ 35\xi ^{6}\left( \cosh (\xi ^{2})+\cos (\xi
^{2})-2\cosh \left( \frac{\xi ^{2}}{\sqrt{2}}\right) \cos \left( \frac{\xi
^{2}}{\sqrt{2}}\right) \right) \right. \right.  \nonumber \\
&&+280\xi ^{4}\left( \sinh (\xi ^{2})+\sin (\xi ^{2})-\sqrt{2}\sinh \left( 
\frac{\xi ^{2}}{\sqrt{2}}\right) \cos \left( \frac{\xi ^{2}}{\sqrt{2}}%
\right) \right.  \nonumber \\
&&\left. -\sqrt{2}\cosh \left( \frac{\xi ^{2}}{\sqrt{2}}\right) \sin \left( 
\frac{\xi ^{2}}{\sqrt{2}}\right) \right)  \nonumber \\
&&+622\xi ^{2}\left( \cosh (\xi ^{2})-\cos (\xi ^{2})-2\sinh \left( \frac{%
\xi ^{2}}{\sqrt{2}}\right) \sin \left( \frac{\xi ^{2}}{\sqrt{2}}\right)
\right)  \nonumber \\
&&+420\left( \sinh (\xi ^{2})-\sin (\xi ^{2})+\sqrt{2}\sinh \left( \frac{\xi
^{2}}{\sqrt{2}}\right) \cos \left( \frac{\xi ^{2}}{\sqrt{2}}\right) \right. 
\nonumber \\
&&\left. \left. \left. -\sqrt{2}\cosh \left( \frac{\xi ^{2}}{\sqrt{2}}%
\right) \sin \left( \frac{\xi ^{2}}{\sqrt{2}}\right) \right) \right]
\right\} .
\end{eqnarray}
Figure 4a plots $S_{\varphi ,8}^{(K=4)}$ as a function of $\varphi $ for
three selected values of $|\xi |.$ From the figure, it is clear that maximal
squeezing occurs along the four directions $\varphi =\varphi _{sq,1}=\pi /8$ 
$(9\pi /8),$ $\varphi =\varphi _{sq,2}=3\pi /8$ $(11\pi /8),$ $\varphi
=\varphi _{sq,3}=5\pi /8$ $(13\pi /8)$ and $\varphi =\varphi _{sq,4}=7\pi /8$
$(15\pi /8),$ whereas for maximal stretching $\varphi =\varphi _{st,1}=0$ $%
(\pi ),$ $\varphi =\varphi _{st,2}=\pi /4$ $(5\pi /4),$ $\varphi =\varphi
_{st,3}=\pi /2$ $(3\pi /2)$ and $\varphi =\varphi _{st,4}=3\pi /4$ $(7\pi
/4).$ The values of $|\xi |$ for which squeezing appears lie within the
interval $[0,0.823267]$ and a maximal squeezing is reached at $|\xi
|=0.754939.$ The coexistence of the four squeezing directions can also be
viewed by a polar plot (Fig. 4b) in which $S_{\varphi ,8}^{(K=4)}$ is drawn
in dependence on $\varphi $ for $|\xi |=0.754939.$

Expressions of amplitude squeezing for higher values of $K$ and $N$ have
also been obtained which are much more cumbersome. The general results
inferred from their detailed analysis will be drawn below. Before doing so
let us address a delicate issue. By construction, fan-states display the
same property along directions separated by a multiple of $\pi /K.$ As $K$
is even, two quadratures are always among the equivalent directions.
Therefore, if squeezing appears along a direction $\varphi _{sq},$ then the
one that is perpendicular to it, $\varphi _{sq+\pi /2},$ is also equally
squeezed. The product of the two quadratures $\left\langle (\Delta
X_{\varphi _{sq}})^{N}\right\rangle _{F}\left\langle (\Delta X_{\varphi
_{sq}+\pi /2})^{N}\right\rangle _{F}$ is obviously less than $\left\langle
(\Delta X_{\varphi _{sq}})^{N}\right\rangle _{CS}\left\langle (\Delta
X_{\varphi _{sq}+\pi /2})^{N}\right\rangle _{CS}=\left( \left( N-1\right) !!/%
\sqrt{2^{N}}\right) ^{2}=R_{N}^{2}.$ Does this mean that the coherent state
does not minimize the uncertainty product as was questioned in \cite{lynch86}
? It was partly answered in \cite{hm} where the authors opened the idea that
in such situations a pair of quadratures are no longer the right canonical
conjugates meant in the Heisenberg inequality and suggested choosing two
field components that differ in phase by less than $\pi /2$ as more
appropriate as canonical conjugates. In our fan-states the two amplitude
components, one along a squeezing direction, the other along the next
stretching direction, are best candidates for the right conjugates. Yet,
whether or not is the CS a minimum uncertainty state (MUS) remained not
clarified. In our opinion, the uncertainty is best evaluated by its area
rather than product. The uncertainty area, $A_{N}^{(K)},$ is given by 
\begin{equation}
A_{N}^{(K)}=\frac{1}{2}\int_{0}^{2\pi }d\varphi \left\langle (\Delta
X_{\varphi })^{N}\right\rangle _{F}^{2}=\pi \left\{ R_{N}^{2}+\left(
2R_{N}+X_{N}^{(K)}\right) X_{N}^{(K)}+\frac{1}{2}\sum_{p=1}^{[N/(2K)]}\left(
Y_{N}^{(K)}(p)\right) ^{2}\right\}
\end{equation}
where $[x]$ equals the integer part of $x,$%
\begin{equation}
X_{N}^{(K)}=\frac{N!}{2^{N}}\sum_{m=1}^{N/2}\frac{2^{m}\left\langle
a^{+m}a^{m}\right\rangle _{F}}{(m!)^{2}\left( N/2-m\right) !}  \label{X}
\end{equation}
which is always positive and 
\begin{equation}
Y_{N}^{(K)}(p)=\frac{2^{pK}N!}{2^{N-1}}\sum_{m=0}^{[N/2-pK]}\frac{%
2^{m}\left\langle a^{+m}a^{m+2pK}\right\rangle _{F}}{m!(m+2pK)!(N/2-m-pK)!}.
\label{Y}
\end{equation}
Because $\pi R_{N}^{2}$ is the circle area of uncertainty in the CS, $%
Y_{N}^{(K)}(p)$ is squared and $X_{N}^{(K)}>0$ the uncertainty in the
fan-states is always greater than that in the coherent state. Hence, the
coherent state remains a right MUS as far as the fan-states are concerned.
Problems arising from more subtle questions such as how the fan-states show
up with the Schr\"{o}dinger-Robertson uncertainty relation, whether the
coherent state is an intelligent state for higher-order moments, etc. are
certainly beyond this letter scope. On this topic, recommended is a recent
paper \cite{tri} which contains detailed discussions as well as a good list
of references.

\noindent {\bf 4. Conclusion}

In conclusion, the fan-state has been constructed as a superposition state
of the symmetric-even K-quantum nonlinear coherent states which in turn are
superposition states of the more elementary single-quantum nonlinear
coherent states. Such a secondary superposition provides the fan-state with
the symmetry leading to multi-directional squeezing. In these fan-states
higher-order amplitude squeezing has been calculated analytically for a wide
set of $K$ and $N$. The general results are summarized as follows. For a
fixed $K$ an amplitude component cannot be squeezed at all in orders $N$
less than $2K.$ The lowest order in which squeezing may appear is $N_{\min
}=2K.$ Squeezing may also occur in an arbitrary (even) order greater than $%
N_{\min }.$ That is, given $K$ the orders in which amplitude squeezing can
be observed is $N_{sq}=2(K+n)$ with $n=0,1,2,...$ For fixed $K$ and $N_{sq},$
squeezing, whenever it exists, is $K$-directional, i.e. it occurs equally
along $K$ directions determined by the angles $\varphi =\varphi
_{sq,m}=[(1+2m)\pi ]/(2K)$ with $m=0,1,...,K-1$ while for stretching
directions $\varphi =\varphi _{st,m}=\pi m/K.$ The uncertainty domain in the
complex plane has the shape of a $2K$-winged flower (see, e.g., Fig. 3b) and
the squeeze parameter $S_{\varphi ,N_{sq}}^{(K)}$ itself, in polar
coordinates, shows up as a $4K$-winged flower (see, e.g., Figs. 3c and 4b)
with $2K$ small wings associated with squeezing and $2K$ big wings
associated with stretching. In the fan-state, which prompts higher-order
squeezing, the two quadratures can no longer be served as the right pair of
canonical conjugates. The appropriate ones should be two amplitude
components which are $\pi /(2K)$ dephased from each other (i.e., one
component points along a squeezing direction and the other component points
along the stretching direction next to the squeezing one). Though being
squeezed in more than one direction the uncertainty area associated with the
fan-state is found always larger than that in the coherent state revealing
the relevance of the coherent state as a MUS.

Multi-quantum states associated with higher-order squeezing have been
studied intensively in the literature. For example, linear four-photon
states were investigated in \cite{l4} and squeezing was calculated up to
order $N=8.$ Yet, squeezing was found to occur only in one direction. In
this letter we concentrate on the multi-direction possibility of squeezing
rather than on the nonlinearity brought in by the function $f.$ As was
mentioned above, the number of squeezing directions in fan-states does not
depend on whether $f=1$ or $f\neq 1.$ The formalism has, however, been
formulated explicitly in the evident presence of $f$ and ready for its
actual inclusion. The shape of uncertainty flower would be dramatically
modified (the number of flower wings is preserved for a fixed $K)$ and, even
the possible existence of the fan-states themselves should be analyzed
carefully with each specific type of $f\neq 1$ in dependence on the
parameters involved. This would constitute an intriguing piece of research
and will be done in the future.

\noindent {\bf Acknowledgments}

The author would like to thank the referee for his/her useful comments and
suggestions. This work was supported by the National Center for Theoretical
Sciences (Physics Division), Hsinchu, Taiwan, R.O.C.

\begin{center}
{\Large FIGURE CAPTIONS}
\end{center}

\begin{description}
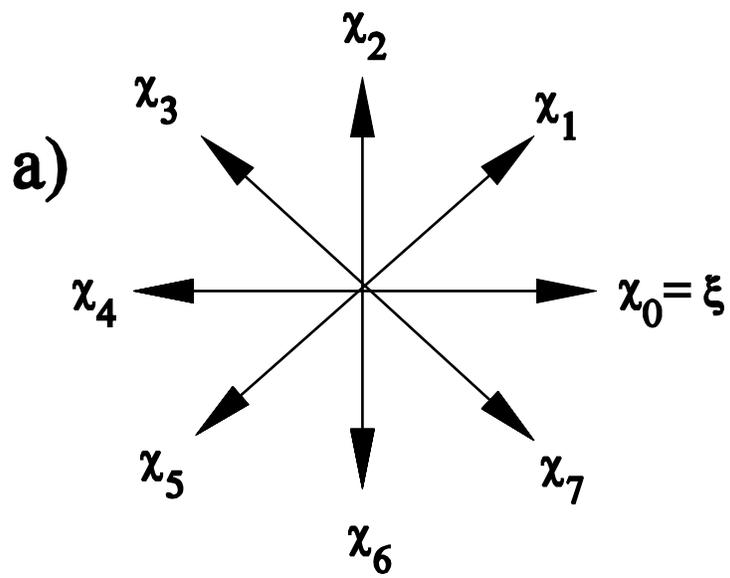
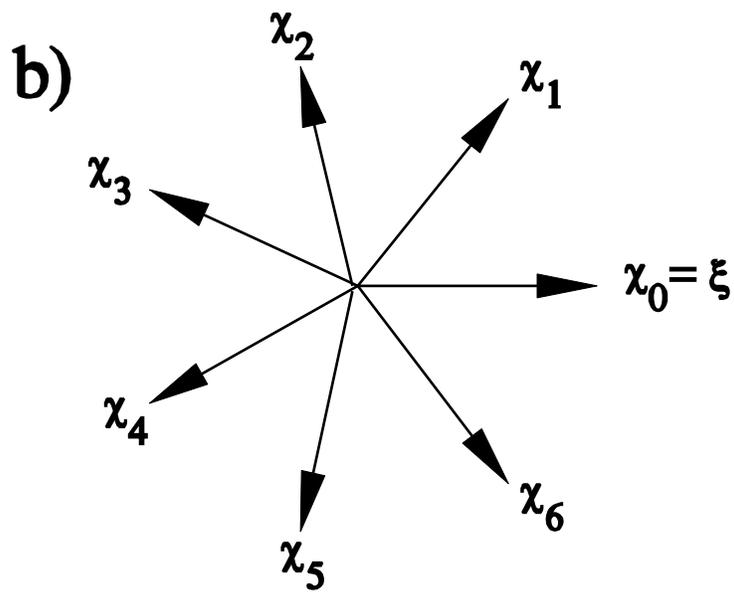
\item[Fig. 1:]  a) Orientation of the $\chi _{l},$ Eq. (\ref{14a}), in the
complex plane in a symmetric-even K-quantum nonlinear coherent state with an
even $K=8.$ b) The same in a symmetric (neither even nor odd) K-quantum
nonlinear coherent state with an odd $K=7.$

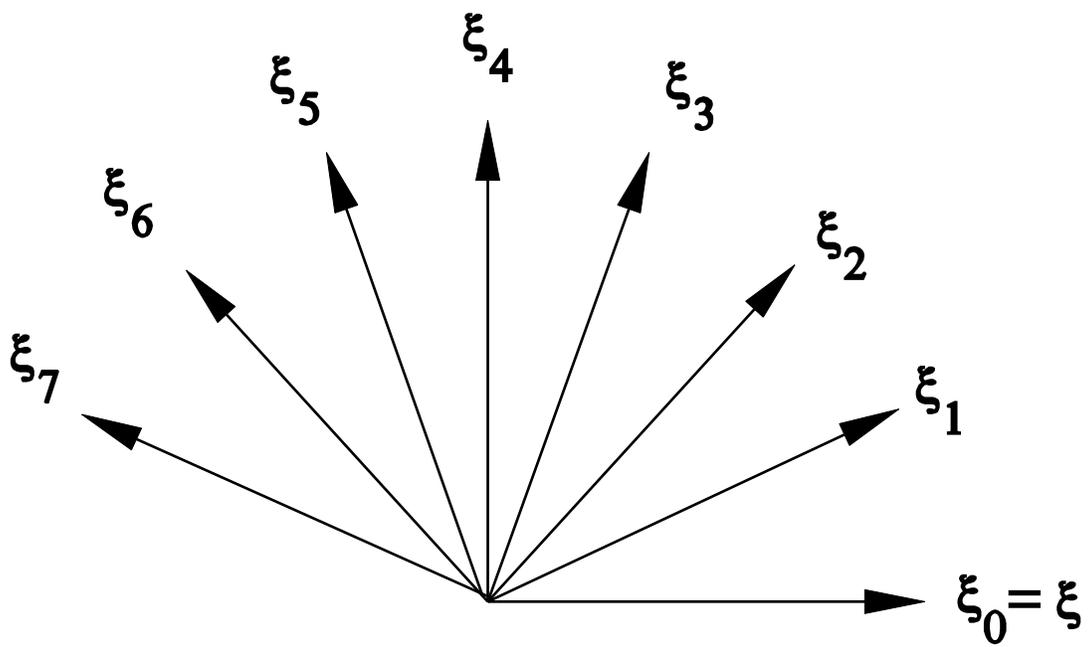
\item[Fig. 2:]  Orientation of the $\xi _{q},$ Eq. (\ref{xiq}), in the
complex plane in a fan-even K-quantum nonlinear coherent state with $K=8.$

\item[Fig. 3:]  a) $S\equiv S_{\varphi ,6}^{(K=2)}$ as a function of $|\xi |$
and $\varphi .$ b) The uncertainty domain in the coherent state (dashed
circle) and in the fan-state (solid four-winged flower) with $K=2$ for $|\xi
|=0.659657$. It is visual that, though squeezed in two directions, the area
bounded by the flower is larger than that bounded by the circle. c) $%
S_{\varphi ,6}^{(K=2)}$ in polar coordinates for the same value of $|\xi |$
as in b). The small wings correspond to squeezing, the big ones to
stretching and the center to the coherent state. The distance from the
center to the farthest point of a big wing is $1.07.$

\item[Fig. 4]  a) $S\equiv S_{\varphi ,8}^{(K=4)}$ as a function $\varphi $
for $|\xi |=0.754939,$ $0.823267$ and $0.85,$ upwards. b) $S_{\varphi
,8}^{(K=4)}$ in polar coordinates for $|\xi |=0.754939.$ The small wings
correspond to squeezing, the big ones to stretching and the center to the
coherent state. The distance from the center to the farthest point of a big
wing is $0.02.$
\end{description}

\end{document}